\documentclass[11pt,twoside]{article}
\usepackage{asp2004}
\usepackage{psfig}
\usepackage{epsf}
\usepackage{graphicx}
\usepackage{lscape}
\def\edcomment#1{\iffalse\marginpar{\raggedright\sl#1\/}\else\relax\fi}
\reversemarginpar

\begin{document}
\title{Supernova search at intermediate \boldmath $z$\\
III. Expansion velocities of the ejecta}
\author{A.~Balastegui~(1), P.~Ruiz-Lapuente~(1), J.~M\'endez~(1,2), G.~Altavilla~(1), M.~Irwin~(3), K.~Schahmaneche~(4), C.~Balland~(5,6), R.~Pain~(4), N.~Walton~(3)}
\affil{1) Department of Astronomy, CER for Astrophysics, Particle
Physics and Cosmology, University of Barcelona, Diagonal 647, E--08028, 
Barcelona, Spain \\
2) Isaac Newton Group of Telescopes, 38700 Santa Cruz de La Palma, 
Islas Canarias, Spain \\
3) Institute of Astronomy, University of Cambridge, Madingley Road, 
Cambridge. CB3 0HA, United Kingdom\\
4) LPNHE-IN2P3-CNRS-Universit\'es Paris 6 et Paris 7, 4 place Jussieu, 
75252 Paris Cedex  05 France\\
5) Institut d'Astrophysique Spatiale, B\^{a}timent 121, Universit\'e
Paris 11, 91405 Orsay Cedex, France\\
6) Universit\'e Paris Sud, IAS-CNAS, B\^{a}timent 121,
Orsay Cedex, France}

\begin{abstract}
We discuss the expansion velocities of different elements in 
the ejecta of the intermediate--z SNe Ia discovered as
 a part of the International Time Programme (ITP)
 project ``$\Omega$ and $\Lambda$ from Supernovae and the Physics of
 Supernova Explosions'' at the European Northern Observatory (ENO). 
The expansion velocities measured for each normal SNIa are found to
be within the typical velocity dispersion for their epoch. Meanwhile,
the subluminous SN 2002lk SiII expansion velocity is significantly
higher than that of SN 1991bg  shortly after maximum. The observed
phase was younger in SN2002lk than in the local subluminous
SNIa SN1991bg.

\end{abstract}

\thispagestyle{plain}

\section{Expansion velocities of the ejecta}

Expansion velocities for CaII (3950\AA) and SiII (6355\AA) lines have been measured for each SNIa, whenever the lines were present and well defined in the spectra. A summary of the results is shown in Table \ref{tabla3}.

The uncertainty in the expansion velocity is measured by square summing
the standard deviation of several $\lambda$ measurements, and adding
to that an error in redshift of $\pm$0.001. The uncertainty in the
epoch, in respect to maximum, when measured with comparison spectra
 is commonly assumed to be approximately of $\pm$2 days \citep{riessepoch}.

   \begin{table}[t]
     \centering
      \caption[]{Expansion velocities for CaII (3950\AA) and SiII (6355\AA) lines.}
         \label{tabla3}
     $$ 
         \setlength\tabcolsep{1pt}
	\begin{tabular}{l c c}
	\hline
	    \hline
            \noalign{\smallskip}
Object & Expansion velocity & ~Expansion velocity \\
name & CaII (km s$^{-1}$) & ~SiII (km s$^{-1}$) \\

	\noalign{\smallskip}
            \hline
            \noalign{\smallskip}

SN 2002li  & 17200$\pm$500 &	\\
SN 2002lj  & 12300$\pm$300 & 8900$\pm$300 \\
SN 2002lp  & 13800$\pm$900 & 10400$\pm$400 \\
SN 2002lq  & 18900$\pm$1400 &	\\
SN 2002lr  & 12800$\pm$300 & 8600$\pm$400 \\
SN 2002lk  & 		& 14500$\pm$500 \\

            \hline
            \hline
         \end{tabular}
     $$ 
\end{table}

\begin{figure}
    \plotfiddle{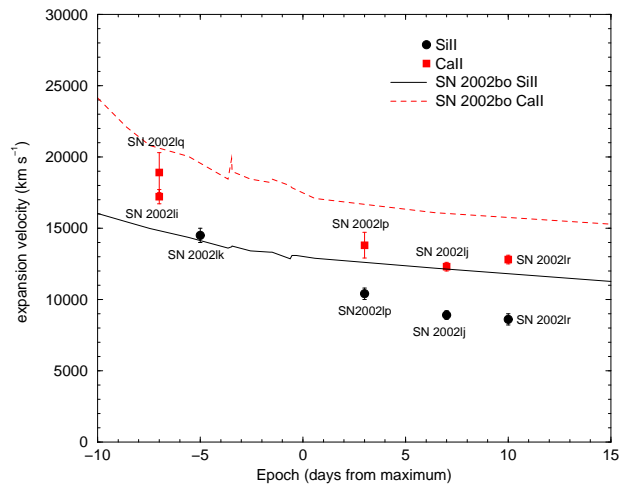}{5.5cm}{-90}{40}{40}{-140}{220}
      \caption{Expansion velocities for CaII (3950\AA) and SiII (6355\AA) lines as a function of the epoch from maximum. It is also shown, as a reference, the expansion velocities for SN 2002bo.}
         \label{vels}
   \end{figure}

Figure \ref{vels} shows the expansion velocity for the CaII 3950\AA and
 SiII 6355\AA lines versus epoch from maximum, plotted for all SNe
 Ia. Reference evolution tracks are also included corresponding to the
 recently well followed nearby SN 2002bo \citep{benettibo}. All
 points lie within the typical dispersion of expansion velocities for
 normal SNe Ia. On the other hand, the subluminous SN Ia SN 2002lk
 has a SiII expansion velocity (14500 km s$^{-1}$) significantly higher
 than the velocity measured for SN 1991bg (9900 km s$^{-1}$)
 near maximum. However, the observed phase of SN 2002lk is
 earlier than the one of
 SN1991bg.\footnote{This research was carried out within the International 
Time Programme
"Omega and Lambda from Supernovae and the Physics of Supernova
Explosions" at ENO.
The acknowledgement is written only in this poster
paper for space limitations,
but also implicit in the other previous two poster papers on ITP
 results. We are grateful to support by grants 
ESP20014642--E, 
 UNI/2120/2002 and HPRN--CT--20002-00303.}

\end{document}